\documentclass{article}
\usepackage{graphics}
\usepackage{amssymb, amsmath}
\title{Single lens telescope}

\begin{document}
\maketitle

\begin{center}
\begin{minipage}{0.5\textwidth}
Rafael Guillermo Gonz\'alez-Acu\~na ${^1,^*}$\\  
H\'ector Alejandro Chaparro-Romo ${^2}$\\
Julio Cesar Guti\'errez-Vega ${^1}$.\\
\end{minipage}\end{center}

\noindent{$^1$} Photonics and Mathematical Optics Group, Tecnol\'ogico de Monterrey,\\ Monterrey 64849, M\'exico.\\[2mm]
{$^2$}  Independent  researcher.\\

{\bf email} rafael123.90@hotmail.com

\begin{abstract}
We present a singlet lens that behaves like a telescope, it expands light rays and at the incoming and outgoing the rays are collimated. Therefore we called singlet lens telescope. The analytical model adapts the second surface when the first surface is given, in order that at the output the rays are collimated. We test the model using ray tracing tracking for several input surfaces and the results are quite satisfactory.
\end{abstract}

\section{Introduction}
A telescope is an optical device which has two or more lenses. At the entrance and exit of the telescope the rays are collimated along the optical axis. The difference is that at the exit, the rays have been concentrated. There are many types of telescopes with different designs and shapes, but all of them have at least two or more lenses \cite{ terebizh2011new,
korsch1972closed, pierre2003design, hanany2002comparison, saha2007diffraction}. 

Historically, the invention of the telescope is attributed to Hans Lippershey, a german lens maker, in 1608 \cite{van1975historical}, but recent research suggests that the first inventor was Juan Roget in 1590 \cite{lopez1979ciencia,santo2017refracted}. The telescope was made popular by Galileo Galilei, when he observed the moon and the rings of Jupiter 409 years ago with his telescope \cite{galilei2016sidereus,sirtorihieronymi} 
Since its early era, no refractive telescope design has appeared with less than two lenses or optical devices \cite{romano2016geometric}.

Recently, lenses free spherical aberration have been discussed \cite{ valencia2015singlet, gonzalez2019freeform, valencia2017catadioptric,gonzalez2018generalization,gonzalez2018general}. One can use two or more of these lenses to easily design a telescope.
 
In this paper, we present a new thick lens design that serves as a telescope, that is, at the entrance and at the exit it has collimated rays along the optical axis. The presented model does not use paraxial approximation and is free of spherical aberration.  The derivation is free of numerical approaches and optimization processes. As far as the authors know, the singlet telescope presented in this paper has not been reported before in the optics literature.

\begin{figure}[h!]
\centering
\resizebox{100mm}{70mm}{\includegraphics{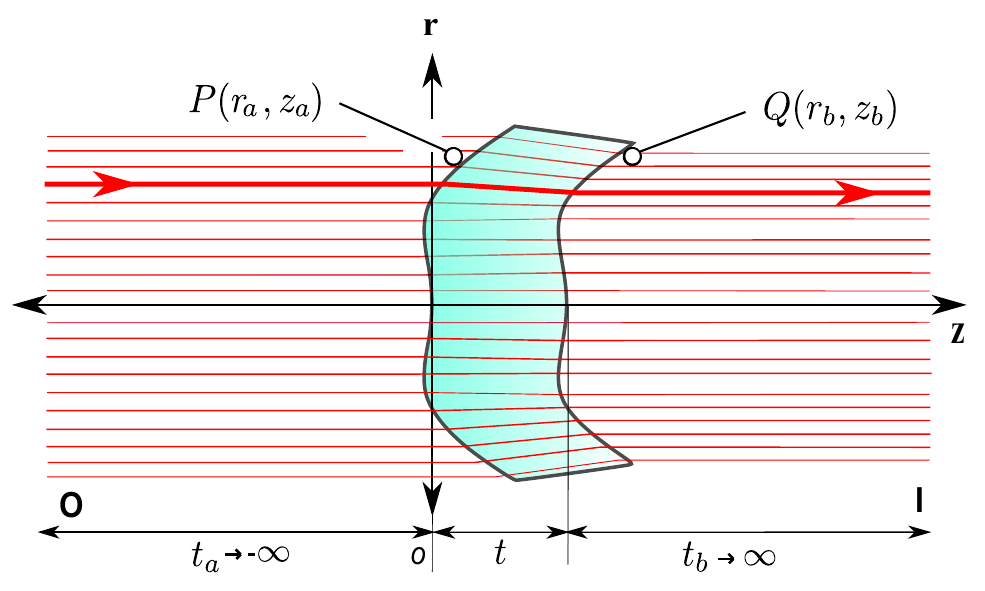}}
\caption{Diagram of a singlet lens telescope. The first surface is given by $(r_a,z_a)$, the second surface is given by $(r_b,z_b)$. The distance between the first surface to the object is $t_a\to-\infty$, the thickness at the center of the lens is $t$ and the distance between the second surface to the~image~is~$t_b\to\infty$.}
\label{portada}
\end{figure}

\section{Mathematical model}
It is well known that the only surface that receives collimated rays and delivers collimated rays is a flat surface but in this case there is no beam expansion. This brings up the question: Given an input surface, hat second surface collimates  rays at exit?

To answer this question, let us first to introduce the following assumptions: the telescopic-lens has a constant refraction index $n$ and is circularly symmetric, it is surrounded by air, and at the optical axis the  telescopic-lens  has  thickness $t$. Let $(r_a,z_a)$ and $(r_b,z_b)$ be the first and second surfaces, respectively, where $z_a$, $z_b$, and $r_b$ are functions of the independent radial coordinate $r_a$ of the input surface. So, the basic question is: given $(r_a,z_a)$ how should be $(r_b,z_b)$ to get a singlet lens telescope?

\begin{figure}
\centering
\resizebox{70mm}{40mm}{\includegraphics{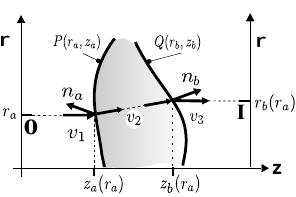}}
\caption{ In the left side of the image, it can be seen three unitary vectors: $\boldsymbol{v}_1$  is the unitary vector of the  incident ray,  $\boldsymbol{v}_2$ is unitary vector of the refracted ray and $\boldsymbol{n}_a$ is the normal vector of the first surface. In the right side of the image, it can be seen three unitary vectors of the second surface:  $\boldsymbol{v}_2$ is the unitary vector of the incited ray and $\boldsymbol{v}_3$ is unitary vector of the refracted ray  and $\boldsymbol{n}_b$ is the normal vector of the second surface.}
\label{vectores}
\end{figure}

The input field impinging on the singlet lens is a monochromatic plane wave traveling into positive $z$ direction. Since the size of the singlet lens is much larger than the wavelength of the light, a ray optics representation may be applied to solve this problem. In this approach, the input field is characterized by an uniform bundle of parallel rays. As shown in Fig. \ref{portada}, the telescope condition requires that all input rays in the same meridional plane emerge parallel from the singlet lens respect to the $z$-axis.

Let  start with the first fundamental equation of our system, the Snell's law at the first surface $a$,
\begin{eqnarray}
\hspace{-1cm}
\label{eq:sin1}
n=\frac{\sin(\theta_{ai})}{\sin(\theta_{ar})}=\frac{\sqrt{1-\cos^2(\theta_{ai})}}{\sqrt{1-\cos^2(\theta_{ar})}}=
\frac{\sqrt{1-(\boldsymbol{\vec{v}}_1\cdot\boldsymbol{\vec{n}}_a )^2}}{\sqrt{1-(\boldsymbol{\vec{v}}_2\cdot\boldsymbol{\vec{n}}_a )^2}},
\label{snell01}
\end{eqnarray}
where $\theta_{ai}$ is the angle of the incident ray respect to the normal $\boldsymbol{\vec{n}}_a$, $\theta_{ar}$ is the refracted ray angle respect to the normal $\boldsymbol{\vec{n}}_a$. Let  $\boldsymbol{\vec{v}}_1$  be
unitary vector of the  incident ray and let $\boldsymbol{\vec{v}}_2$ be unitary vector of the refracted ray,  see Fig. \ref{vectores}.

The unitary vectors are given by,
\begin{eqnarray}
\label{vec1}
\boldsymbol{\vec{v}}_1=[0,1],\quad
\boldsymbol{\vec{v}}_2=\displaystyle{\frac{[r_b-r_a,z_b-z_a]}{\sqrt{(r_b-r_a)^2+(z_b-z_a)^2}}},\quad
\boldsymbol{\vec{n}}_a=\displaystyle{\frac{[z_a',-1]}{\sqrt{1+(z_a')^2}}}.
\end{eqnarray}
Replacing  Eq. \ref{vec1} into Eq. \ref{snell01} and after some algebraic simplifications, we get
\begin{eqnarray}
\begin{cases}
\displaystyle{\frac{r_b-r_a}{\sqrt{(r_b-r_a)^2+(z_b-z_a)^2}}=-\frac{z_a' \left( \sqrt{\left(n^2-1\right) z_a'^2+n^2 }-1\right)}{n \left(z_a'^2+1\right)}}\equiv\chi,\\
\displaystyle{\frac{z_b-z_a}{\sqrt{(r_b-r_a)^2+(z_b-z_a)^2}}=\frac{\sqrt{\left(n^2-1\right)z_a'^2+n^2}}{n(z_a'^2+1)}+\frac{z_a'^2}{n z_a'^2+n}}\equiv\zeta.
\end{cases}
\label{snell1}
\end{eqnarray}

Now let us apply the Fermat's principle to the problem.
For a spherical aberration-free telescopic-lens, the optical path of any non-central ray  must be equal to the optical path of the axial ray. For derivation purposes, it is convenient for the moment to consider that the rays come from a finite object located on the optical negative $z$-axis at $z=-t_a$,  and focuses at an image image point located at a finite distance $z= t+t_b$. Therefore, for a finite object and a finite image, both located on the optical axis, we have following equality for the optical paths:
\begin{eqnarray}
\label{fer0}
-\text{sgn}(t_a)\sqrt{r_a^2+(z_a-t_a)^2}+n\sqrt{(r_b-r_a)^2+(z_b-z_a)^2}\\
+\text{sgn}(t_b)\sqrt{r_b^2+(z_b-t-t_b)^2}=-t_a+nt+t_b, \nonumber
\end{eqnarray}
where $t_a$ is the distance from the object to the first surface and $t_b$ is the distance from the second surface to the image. 

Now to recover the parallelism of the input and output rays, we will let the object point to tend to minus infinity, i.e. $t_a\rightarrow-\infty$, and the image point to tend plus infinity $t_b\to +\infty$. After applying both limits in Eq. (\ref{fer0}), we obtain
\begin{eqnarray}
n=\displaystyle{\frac{t+z_a-z_b }{t-\sqrt{(r_b-r_a)^2+(z_b-z_a)^2}}},
\label{fer1}
\end{eqnarray}
We have two algebraic relations, Eqs. \ref{snell1} and \ref{fer1}, and two unknowns $z_b$, $r_b$. The solution of this system is given by
\begin{equation}
\label{sol}
z_b=\displaystyle{\frac{n z_a-(t-nt-z_a) \zeta}{n-\zeta} },\quad
r_b=\displaystyle{\frac{\chi(z_b-z_a)}{\zeta}+r_a}.
\end{equation}
Equations (\ref{sol}) are the most important result presented in this paper. In fact they describe analytically the shape of the second surface $z_b(r_b)$ given the first surface $z_a(r_a)$ for the lens parameters $(n,t)$. In the process of derivation it was not needed to use any numerical approach nor an iterative optimization process. These expressions may look cumbersome, but it is quite remarkable  that the output shape may be expressed in closed-form for an almost arbitrary input surface.  

We remark that the first surface must be rotationally symmetric, its normal should be parallel to the optical axis at the origin, and the rays travelling inside the lens should not cross each other. Let us discuss the last condition in more detail. From a topological point of view, the singlet lens is an homogeneous optical element, i.e. the input and output surfaces are simple connected sets on $\mathbb{R}^2$ that can be defined as
\begin{equation}
\Psi_a=\{(r_a,z_a)\in \mathbb{R}^2| z_a<z_b  \},\hspace{1cm}
\Psi_b=\{(r_b,z_b)\in \mathbb{R}^2| z_b>z_a  \},
\end{equation}
where $\Psi_a$ and $\Psi_b$ are homeomorphic. Topologically speaking  it  means that both surfaces are topologically equivalent. Thus, there exists a continuous and bijective function $f$ such that $f:\Psi_a \to \Psi_b$, and whose inverse $f^{-1}$ is continuous. 

There are many functions $f$ that map both sets, but there is only one  is physically valid function given by Eqs. (\ref{sol}).  Now, since $f$ is continuous it means that $f$ maps open balls from $\Psi_a$ to $\Psi_b$, then the ray neighborhoods are preserved. Therefore, the validity of Eqs. (\ref{sol}) also requires that the rays do not intersect each other inside the lens, as we mentioned earlier. In the case when the rays cross each other, $\Psi_b$ overlaps itself leaving from being homeomorphic with respect to $\Psi_a$, and the vicinity of the neighborhoods are not preserved, therefore we do not have a homogeneous optical element. By contradiction we need that the first surface is such that the rays inside the lens do not cross each other.

\section{Illustrative  examples}
In the following figures we plot some designs of the telescopic-lenses given by Eq. (\ref{sol}) and their corresponding the ray traces. We include the design specifications in the caption of the figures.

\subsection{Robustness} 

\begin{figure}[h]
\centering
\resizebox{1\textwidth}{35mm}{\includegraphics{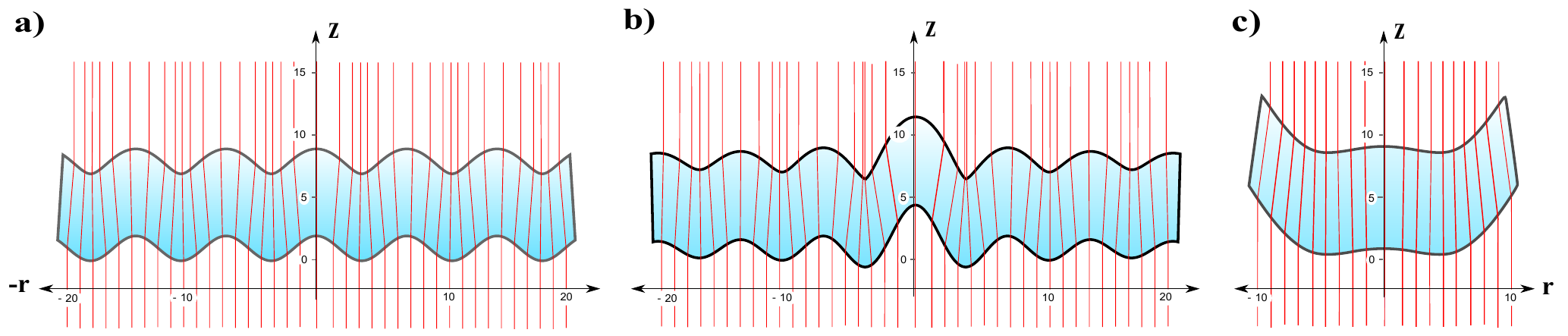}}
\caption{The  configuration case $a)$ is $t=7$, $n=1.1$, 
$z_a=\cos(r_a)$.  The  configuration case $b)$ is $t=7$, $n=1.1$, 
$z_a=3.5J_0(r_a)$.  The  configuration case $c)$ is
$t=8$, $n=1.1$, 
$z_a=r_a^2/18+\cos(r_a/2)-1$. }
\label{f}
\end{figure}

In Fig. \ref{f} we plot several sagittas as first surfaces in order to illustrative the robustness and versatility of Eq. (\ref{sol}). In case $(a)$ the first surface is given by a cosine $z_a=\cos(r_a)$, the spatial frequency of the input function modulating the input surface can be increased until the limit when the rays traversing inside the glass intersect each other. It can be seen by the density of rays that the lens presented in case $(a)$ transforms a uniform incoming field into a non-uniform field at the exit.

In case $(b)$ the sagitta $z_a=3.5J_0(r_a)$ is a Bessel function of order zero. Like in case $(a)$ the first surface is smooth enough that the rays inside do not cross each other. 

In subplot $(c)$  we consider an input surface given by the superposition of a parabola and the cosine function. In this case, the radius of the telescope $R_{max}$ is finite and can be determined by the intersection of the input and output curves, i.e. $z_b(R_{max})=z_a(R_{max})$. Unfortunately, it seems that there is not a close-form analytic expression for $R_{max}$, but it can be calculated numerically finding the cross point of both curves. It is important to mention that if  $R_{max}$ increases as the thickness $t$ increases.  

The shapes presented in figure \ref{f} are just some illustrative examples of the many examples computed. The general formula still giving a correct second surface in order to get a singlet lens free of spherical aberration.

\subsection{Variation of the refraction index}

\begin{figure}[h]
\centering
\resizebox{1\textwidth}{60mm}{\includegraphics{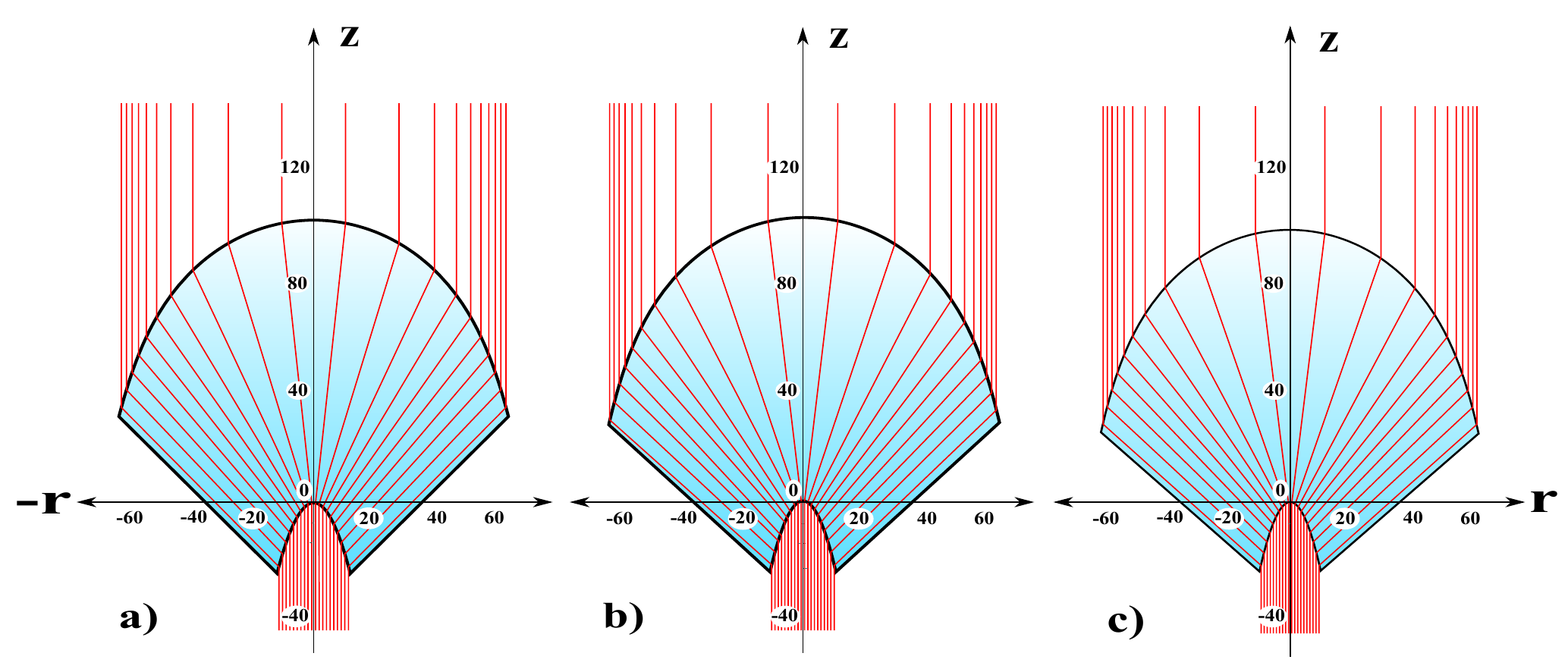}}
\caption{Diagram single lens telescope, $d_l$  is the diameter of the lens and $t$ is the thickens of the lens.}\label{e}
\end{figure}

 In this section we keep constant all input parameters $(z_a,t)$ while we vary the index of refraction $n$ for a singlet-telescope with thickness $t=10$ cm. We use as a first surface the parabola $z_a=-r_a^2/4$. The ray tracing is plotted in Fig. \ref{e} for the refractive indices $n=1.5$,  $n=1.7$ and $n=1.9$. From the figure it is clear that the rays at the input and the output are collimated.

The transverse beam expansion of the singlet can be defined by the ratio of the widths of the output and the input beams, i.e.
\begin{equation}
\text{E}=w_b/w_a, 
\end{equation}
where $w_b$ is the the width of the exit beam and $w_a$  is the the width of the input beam.

For the case $(a)$ E = 4.45 , for $(b)$ E = 5.36, and for $(c)$ E = 5.81. For the three cases the beam expansion is remarkably high taking into account that we are just using one single lens rather than a system of several singlets. The device has $t=10$ cm of central thickens, thus someone can hold it in his o her hands easily. The beam expansion is approximately proportional to the index of refraction. Also the beam expansion is proportional to the central thickness. Unfortunately we do not have an analytical close form to express this relation but it may be computed numerically.

We have tested a large variety of input surfaces exhibiting concave and convex shapes and different spatial variations. In all cases Eq. (\ref{sol}) gave the correct and expected behavior provided that the rays traveling inside the telescope do not self-intersect.

Please notice that in the single lens telescopes proposed in figure (\ref{e}) the first surface is convex, but if we take the first surface as convex spherical the model will compute a second surface such as the exit rays are collimated, but it is important to remark that we do no control if the second surface will be convex or concave and also we do not control the beam expansion of the second surface and since the solution is unique there is only one shape that satisfies that the rays in the input/output  be collimated. The most natural case, when first surface is a spherical concave is not presented since its poor beam expansion.

Also it is important to remark that the presented diagrams and mathematical analysis describes  optical elements which expands a collimated beam into a larger diameter collimated beam. This is the
opposite behavior of a telescope which reduces the diameter of the input beam thereby magnifying the image seen through an eyepiece.
We are actually describing a beam expander which is essentially a telescope
flipped backwards, but since the system is reversible and the rays in the input/output are collimated we just can flipped backwards the diagrams presented a we will get the singlet lens telescope.

\subsection{Accuracy}
To valid the Accuracy of Eq. (\ref{sol}) we compare the vectors $\boldsymbol{v}_3$ and $\boldsymbol{v}_3^\dagger$. The vector $\boldsymbol{v}_3$ comes from the image to the second surface, and $\boldsymbol{v}_3^\dagger$ is computed using the Snell' law with the normal vector  $\boldsymbol{n}_b$ of the second surface and the unitary vector $\boldsymbol{v}_2$. Therefore, $\boldsymbol{n}_b$, $\boldsymbol{v}_3$ , $\boldsymbol{v}_3^\dagger$  are given by

\begin{equation}
\left\{\begin{array}{l}
\boldsymbol{n}_b=\displaystyle{\frac{[z_b',-r_b']}{\sqrt{z_b'^2+ r_b'^2}}},\quad
\boldsymbol{v}_3=\displaystyle{[0,1]},\quad\\[5mm]
\boldsymbol{v}_3^\dagger=\displaystyle{\displaystyle{n[\boldsymbol{n}_b\wedge (-\boldsymbol{n}_b \wedge \boldsymbol{v}_2)]-\boldsymbol{n}_b\sqrt{1-n^2(\boldsymbol{n}_b \wedge \boldsymbol{v}_2)\cdot(\boldsymbol{n}_b \wedge \boldsymbol{v}_2)}}}  \,,
\end{array}\right.
\end{equation}
where $\wedge$ is the wedge product of Cliford's algebra \cite{hestenes2012clifford}.
The percentage accuracy $A$ of ray measures how close it ends in the image position, thus $E$ is defined by
\begin{equation}
A=100\%-\left|\frac{\boldsymbol{v}_3^\dagger-\boldsymbol{v}_3}{\boldsymbol{v}_3}\right|\times 100\%.
\end{equation}
We compute the accuracy for 550 rays for all the examples presented in the paper and the  average of all the examples is $99.99999999999986`\%\approx100\%$. We believe that the error is not exactly zero because when the equations are evaluated we get computational errors such as truncation.

Finally, we remark that for all examples the singlets are free of
spherical aberration even when the width of the incident rays is very
large.

\section{Conclusions}
In conclusion, it is possible to design a singlet lens with arbitrary input surface to perform the action of a telescope. We have determined a close-form analytic expression of the output surface given the shape of the input surface.  We test the model for several meaningful cases and in all the examples the solution gave the expected results. Also we validated numerically the functionality of our results and found that the accuracy of the system is practically 100 \%.

The presented design has countless applications in science, industry, and everyday life, since Eq. (\ref{sol}) is the general analytic formula of a telescope/beam expander. Just mention a few, the beam expander used in research laboratories may be manufactured with a single lens instead of an array of optical devices. It can be used as telescope with the simplest design and configuration, and also can be used in binoculars, field glasses, a monocular or a spotting scope. % In appendix A we have included the program code in Mathematica language to plot the single lens telescope.

Finally we want to remark that since invention of the telescope  approximately in 1590, no  telescope of a singlet lens has been presented. A 429 years old conjecture is satisfactorily solved in this paper.

\bibliographystyle{apalike}
\bibliography{single_lens_telescope}
\end{document}